\documentclass[amsfonts, aip, amssymb, amsmath, reprint]{revtex4-1}

\usepackage{siunitx}
\usepackage[english]{babel}
\usepackage[utf8]{inputenc}
\usepackage[colorinlistoftodos, color=green!40, prependcaption]{todonotes}

\usepackage{graphicx}
\usepackage{dcolumn}
\usepackage{bm}

\usepackage{amsthm}
\usepackage{mathtools}
\usepackage{physics}
\usepackage{xcolor}
\usepackage{graphicx}
\usepackage{adjustbox}
\usepackage{placeins}
\usepackage[T1]{fontenc}
\usepackage{lipsum}
\usepackage{csquotes}
\usepackage[pdftex, pdftitle={Article}, pdfauthor={Author}]{hyperref} 
\begin{document}
\title{Thermal IR detection with nanoelectromechanical silicon nitride trampoline resonators}

\author{Markus Piller}
\affiliation{Institute of Sensor and Actuator Systems, TU Wien, Gusshausstrasse 27-29, 1040 Vienna, Austria.}
\author{Johannes Hiesberger}
\affiliation{Institute of Sensor and Actuator Systems, TU Wien, Gusshausstrasse 27-29, 1040 Vienna, Austria.}
\author{Elisabeth Wistrela}
\affiliation{Institute of Sensor and Actuator Systems, TU Wien, Gusshausstrasse 27-29, 1040 Vienna, Austria.}
\author{Paolo Martini}
\affiliation{Institute of Sensor and Actuator Systems, TU Wien, Gusshausstrasse 27-29, 1040 Vienna, Austria.}
\author{Niklas Luhmann}
\affiliation{Institute of Sensor and Actuator Systems, TU Wien, Gusshausstrasse 27-29, 1040 Vienna, Austria.}
\author{Silvan Schmid}%
 \email{silvan.schmid@tuwien.ac.at}
\affiliation{Institute of Sensor and Actuator Systems, TU Wien, Gusshausstrasse 27-29, 1040 Vienna, Austria.}

\date{\today} 

\begin{abstract}
Nanoelectromechanical (NEMS) resonators are promising uncooled thermal infrared (IR) detectors to overcome existing sensitivity limits. Here, we investigated nanoelectromechanical trampoline resonators made of silicon nitride (SiN) as thermal IR detectors. Trampolines have an enhanced responsivity of more than two orders of magnitude compared to state-of-the-art SiN drums. The characterized NEMS trampoline IR detectors yield a sensitivity in terms of noise equivalent power (NEP) of \SI{7}{\pico \watt \per \sqrt{\hertz}} and a thermal response time as low as 4~ms. The detector area features an impedance-matched metal thin-film absorber with a spectrally flat absorption of  50\% over the entire mid-IR spectral range from \SIrange{1}{25}{\um}.
\end{abstract}

\keywords{NEMS, thermal infrared detector, trampoline resonator, LPCVD silicon nitride}

\maketitle


\section{Introduction}\label{sec:level1}

Thermal detectors for infrared (IR) measurements are essential devices for infrared spectroscopy and thermal imaging \cite{Rogalski2019,Datskos2003detectors,Skidmore}. Due to the flat and broadband spectral response, these detectors are mostly used when measurements have to be performed over a wide spectral range from near-infrared all the way to the far-infrared regime. 
However, state-of-the-art uncooled thermal detectors' sensitivity is still several orders of magnitude below the fundamental detection limit, which is given by power fluctuations of thermal radiation from the detector and its background \cite{Rogalski2019,kruse2004can}.

Thermal detectors absorb the low energy infrared photons and measure the resulting photothermal heating. The temperature increase is usually measured electrically, e.g., via a thermoelectric voltage, resistance change, or pyroelectric voltage. These electrical temperature sensing schemes are typically limited by thermal noise (Johnson noise) \cite{Rogalski2019}. A mechanical IR sensing concept was introduced in the late 60s by Cary Instruments as a promising thermal detector that is not intrinsically limited by the thermal noise limit \cite{Cary1969}. The principle of this concept is a macroscopic tensioned foil resonator that acts as the thermal sensing element. Such macroscopic IR detectors have, to the best of our knowledge, never been successfully implemented at the time. It was only in 2011, when the successful fabrication and characterization of such  nanometer thin tensioned metal and silicon nitride (SiN) foil resonator elements for temperature sensing was first demonstrated~\cite{larsen2011ultrasensitive}. Later in 2013, nanomechanical photothermal detector concepts based on tensioned SiN strings \cite{Yamada2013,Larsen2013} have been introduced. Recently, this concept has been developed further to be used as an infrared detector based on a SiN drum featuring a broadband infrared absorber thin film \cite{Piller2019}. It has been shown these drum resonators can reach an intrinsic sensitivity in the \si{\femto\watt}-regime \cite{Chien2018a}. It has further been shown that these structures enter the radiative heat transfer regime for lateral drum sizes $>\SI{1}{\mm}$ \cite{Piller2020,Sadeghi2020,zhang2020radiative}. Nanomechanical SiN resonators present a promising approach to create thermal IR detectors that can reach the long anticipated photon noise limit.

The same detector concept has been presented with graphene trampolines \cite{Blaikie2019}. Other micro- and nanoelectromechanical (MEMS and NEMS) thermal detector concepts include piezoelectric resonators \cite{Qian2019,Zhang2019,9312194}, torsional paddle resonators \cite{Zhang_Roukes2013,PhysRevApplied.9.024016}, and polymer resonators \cite{adiyan2019shape}.

\begin{figure*}
\centering
\includegraphics[width=0.85\textwidth]{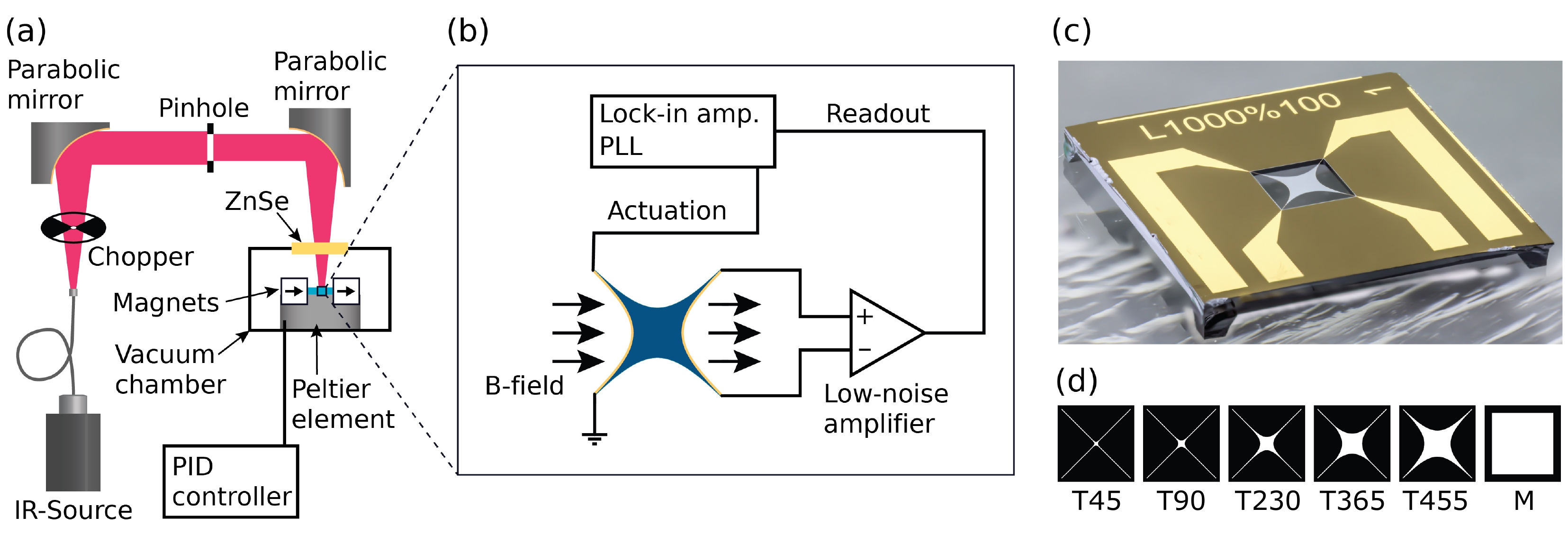}%
\caption{\label{fig:fig_1} (a) Schematic depiction of the measurement setup. (b) Detailed schematic of the magnetomotive transduction scheme. (c) Microscope image of a trampoline resonator sample with gold electrodes (type T455). (d) Depiction of various trampoline resonator types used in this study, reaching from the smallest detection area T45 (\SI{45 x 45}{\um}) to the largest detection area at T455 (\SI{455 x 455}{\um}), and a drum resonator M with a  size of \SI{1000 x 1000}{\um}.}
\end{figure*}

Here, we investigated thermal IR detectors based on nanoelectromechanical SiN trampoline resonators, which already proved exceptional properties in other fields \cite{NortePRL2016,Reinhardt2016,Fischer2019,Chien2020_2}. The experimental setup and the specific SiN trampolines with various detector area sizes that were studied are depicted in Figure~\ref{fig:fig_1}. 

\begin{figure}
\includegraphics[width=0.45\textwidth]{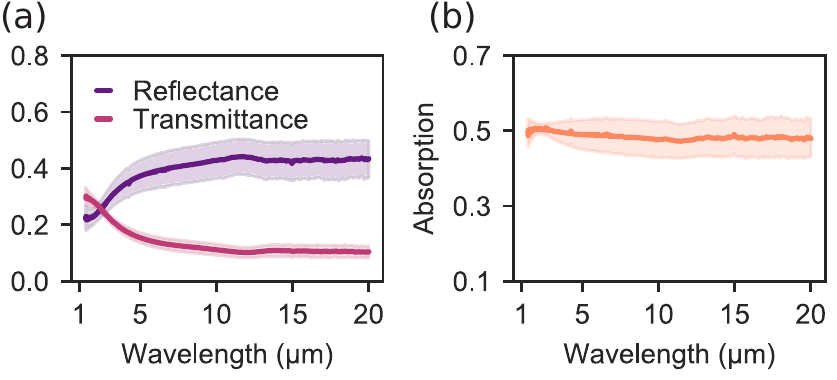}
\caption{\label{fig:absorption} (a) Measured transmittance and reflectance spectra for a \SI{50}{\nano\meter} SiN drum with a \SI{5}{\nano\meter} platinum thin film, (b) and the corresponding absorption spectrum.}
\end{figure}

The NEMS trampoline detectors in this work are transduced by a magnetomotive scheme \cite{Venstra2009,Kough2017} and feature an ultrathin impedance-matched absorber film with a flat spectral response for a wavelength range from \SIrange{1}{25}{\um}, as shown in Figure~\ref{fig:absorption}. The trampoline design results in a responsivity enhancement of more than one order of magnitude over comparable rectangular drum resonators. The responsivity is proportional with the trampoline tether length. The trampolines with the longest tethers measured a noise equivalent power (NEP) of \SI{7}{\pico \watt \per \sqrt{\hertz}} with a thermal response time as low as 4~ms. This is an improvement in sensitivity of two order of magnitude compared to the drum design.

\section{Methods}
\subsection{Experimental setup}
The experimental setup comprises a broadband thermal IR light source (ArcLight IR form Arcoptix) with a spectral range from \SIrange{1}{25}{\um}, an optical chopper (MC2000B and MC1F2 from Thorlabs, Inc.), two parabolic gold mirrors with a variable iris (pinhole) for intensity reduction, and a vacuum detector chamber. The chamber features a sample mount with two permanent magnets and a PID controlled Peltier element to maintain a constant detector temperature of \SI{20}{\degreeCelsius}.

Prior to the characterization of our detector, the incident power of the IR radiation was measured with a reference detector (UM9B-BL-L-D0 from Gentec-Eo). The IR light was passing a fiber with a diameter of \SI{900}{\um}. Using two parabolic mirrors with equal focal length, the optimal IR beam diameter on the detector corresponds to the fiber diameter, resulting in an average power after the zinc selenide (ZnSe) window of $P= \SI{7}{\micro\watt}$.

The magnetomotive transduction was facilitated by permanent magnets that produce a magnetic field of approximately \SI{0.6}{\tesla}. Two gold traces on the SiN trampolines are employed for separated actuation and readout of the nanomechanical motion.
The trampoline's out-of-plane motion induces a voltage along the readout gold trace, which is connected to a differential low-noise voltage preamplifier (SR560 from Stanford Research Systems). A lock-in amplifier with a phase-locked loop module (HF2LI from Zurich Instruments) is used to create an oscillator based on the nanoelectromechanical trampoline resonator, which was operated at the fundamental vibrational mode.

\subsection{Sample fabrication}
The trampoline resonators are made of a low-stress silicon-rich silicon nitride thin film with a thickness of \SI{50}{\nano\meter} that was fabricated by low-pressure chemical vapour deposition. The \SI{5}{\um} wide trampoline tethers are supported by a silicon frame with a thickness of \SI{380}{\um}.
All chips are \SI{5 x 5}{\milli\meter} with a window size of \SI{1 x 1}{\milli\meter}.
The \SI{1}{\um} wide gold electrodes with a thickness of \SI{190}{\nm} and a \SI{10}{\nm} chromium adhesion layer beneath were added by a physical vapour deposition process. 

The SiN trampolines are covered with an IR absorber consisting of a Pt thin film with a thickness of \SI{5}{\nano\meter}. Such an impedance-matched absorber provides optimal absorption of 50\% over the whole spectral range from \SIrange{1}{25}{\um}~\cite{Hilsum1954,Luhmann2020,KENNY1997}, which was confirmed by spectral measurements via Fourier-transform IR spectroscopy (Bruker Tensor 27) equipped with a transmittance and reflectance unit (Bruker A510/Q-T). Figure~\ref{fig:absorption}(a) shows the obtained transmittance $T$ and reflectance $R$ spectra from which the absorption $\alpha$ readily can be calculated as $\alpha = 1 - R - T$.

\section{Results}
\begin{figure}
\includegraphics[width=0.45\textwidth]{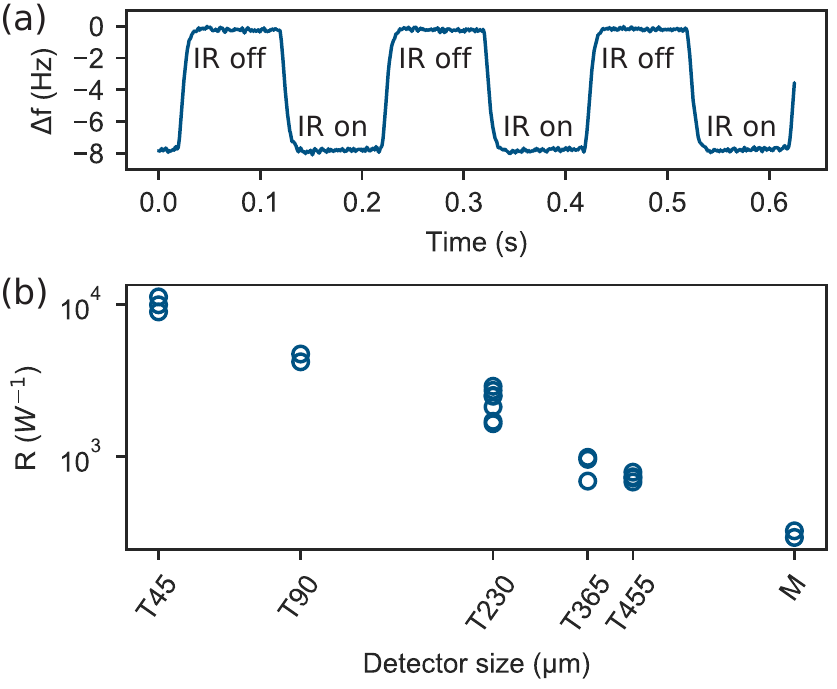}
\caption{\label{fig:responsivity} IR characterization measurements for responsivity. (a) A PLL measured resonance frequency for a chopped IR light at \SI{5}{\hertz} exemplary for a trampoline T45. (b) Relative responsivity obtained for different detector sizes of trampolines. Each data point corresponds to a measurement of an individual sample of the denoted size depicted in Figure~\ref{fig:fig_1}(d).}
\end{figure}

The detection mechanism is based on photothermal detuning of the resonators' resonance frequency. The incident light causes a change in temperature of the resonator and a thermal expansion leading to a reduction of the tensile stress~\cite{Piller2019,Larsen2013,Yamada2013b,Schmid2014d}.
Hence, the responsivity of such a nanoelectromechanical  detector with a detector area $A$ is given by the relative frequency change $\delta f = \Delta f/f_0$ per power of IR light irradiated over the detector area
\begin{equation}
    R = \frac{\delta f}{P_{abs}}.
\end{equation}
The absorbed power $P_{abs}$ is the integrated power over the area of the Gaussian beam profile.
Figure~\ref{fig:responsivity}(a) shows a typical frequency measurement where the infrared light has been turned on and off.  From such time resolved response measurements, the responsivity of each nanoelectromechanical detector was derived, as shown in Figure~\ref{fig:responsivity}(b).  The measured responsivities steadily increase for smaller trampoline detector sizes. Since the frame size is fixed for all detectors to a size of 1~mm, the tether length of the trampolines linearly scales with the detector area. The tethers become longer for a smaller detector area, which causes an improved thermal isolation. The observed enhanced responsivity of trampolines with small detector area can hence be attributed to the increased tether length. The responsivity in the conductive heat transfer regime of the trampolines can be approximated by two crossing strings of length $L$~\cite{Schmid2014d,Schmid2016}
\begin{equation}\label{eq:R}
    R=-\frac{\alpha E L}{32 \kappa \sigma h w},
\end{equation}
with the thermal expansion coefficient $\alpha$, Young's modulus $E$, thermal conductivity $\kappa$, tensile stress $\sigma$, and string cross-section area $h \times w$. $R$ is inversely proportional to $L$ resulting in a linear decrease of $R$ with detector area, as it is clearly observable in Figure~\ref{fig:responsivity}(a), in particular for the smallest trampolines. The measured maximum responsivity of $R=\SI{11000}{\per\watt}$ is more than one order of magnitude below the values obtained with plain silicon nitride drums~\cite{Chien2018a}. According to (\ref{eq:R}), the reason are the Au electrodes that pass over the drum and significantly increase the thermal conductivity. 

\begin{figure}
\includegraphics[width=0.45\textwidth]{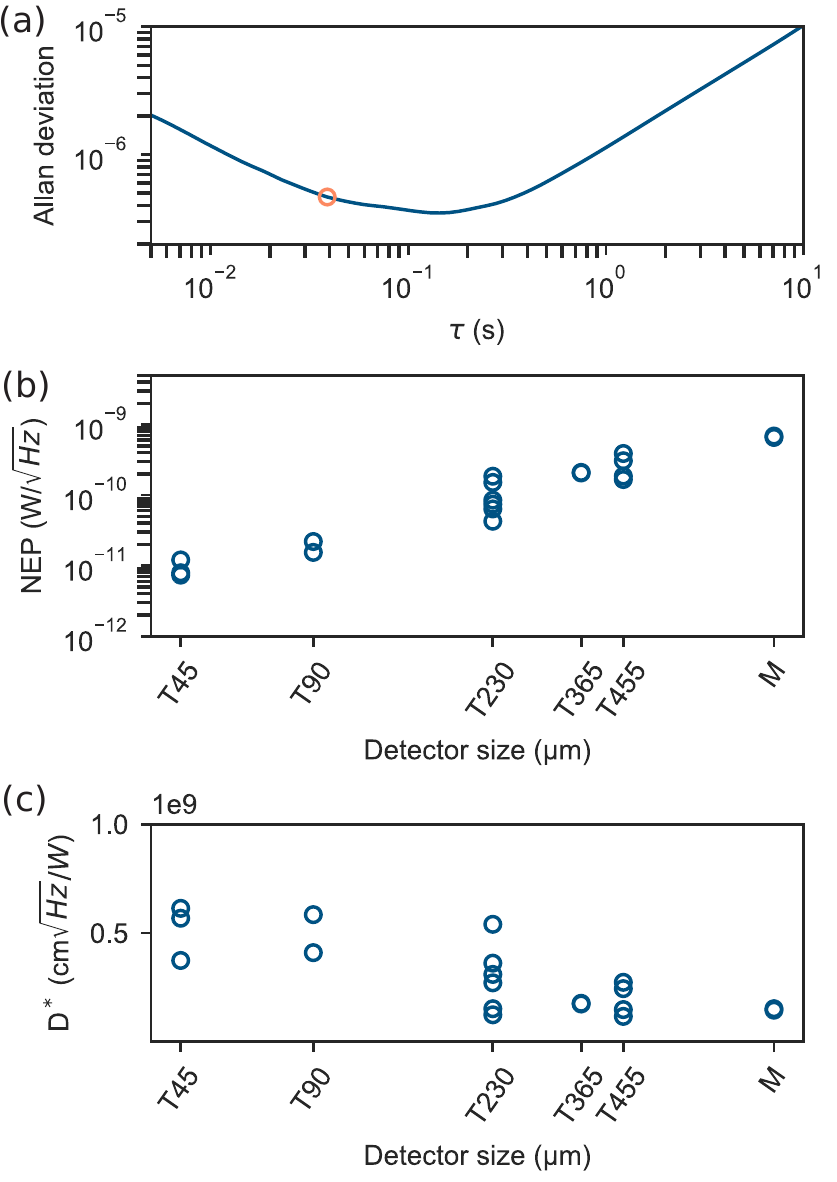}
\caption{\label{fig:NEP} (a) Measured Allan deviation and marked value used for the sensitivity, exemplary for a trampoline T45. (b) NEP and (c) corresponding specific detectivity obtained for various sample sizes, where each data point corresponds to a measured value of an individual sample.}
\end{figure}

Next, the sensitivity is determined in terms of the noise equivalent power (NEP). The NEP of a nanoelectromechanical detector directly scales with the frequency resolution, which was determined through the respective Allan deviation $\sigma_{\text{AD}}$ for a given integration time $\tau$
\begin{equation}
    \label{NEP}
    \text{NEP} = \frac{\sigma_{\mathrm{AD}} \cdot \sqrt{\tau}}{R}.
\end{equation}
The Allan deviation was calculated from frequency recordings over one minute of each nanoelectromechanical resonator. 
An example of the Allan deviation for a T45 sample is shown in Figure \ref{fig:NEP}(a). 
The marker in Figure \ref{fig:NEP}(a) indicates the integration time $\tau=\SI{40}{\milli\second}$ that has been selected to calculate the resulting NEP.

Figure~\ref{fig:NEP}(b) presents the measured NEPs of all nanoelectromechanical trampoline resonators. The NEP improves for smaller trampolines, according to the enhanced responsivity of these structures. 
Compared to the drum resonators (M), the NEP of the trampolines was improved by up to two orders of magnitude. The smallest trampolines showed a NEP of \SI{7}{\pico \watt \per \sqrt{\hertz}}. 

Figure~\ref{fig:NEP}(c) presents the obtained specific detectivity $D^*=\sqrt{A}/\text{NEP}$, which normalizes the sensitivity of a detector with its detection area $A$. $D^*$ is typically used to compare quantum detectors for which noise power is directly proportional to detector size. Noise in thermal detectors does not necessarily follow this trend \cite{Datskos2003}. However, the trampolines' responsivity is proportional to the detector size as discussed above. Hence, the measured specific detectivity values are constant to a good approximation, in particular for the smallest trampolines with the longest tethers. 

\begin{figure}
\includegraphics[width=0.45\textwidth]{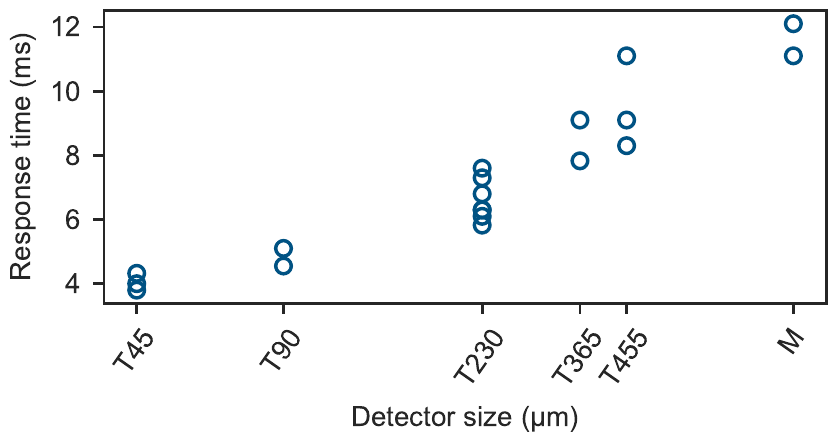}
\caption{\label{fig:response_time} Comparison of the thermal response time analysis between different trampolines and the drum resonator.}
\end{figure}

Figure~\ref{fig:response_time} shows the measured response times, which were obtained from the \textit{90/10 method} \cite{Duraffourg2018} by calculating the rise time from step transition and relating it to a first-order low-pass filter model. The trampolines with the smallest detector size show an improved behavior towards faster response times. The response time of a thermal detector $\tau = C/G$ is given by the ratio of heat capacity $C$ to heat conductance $G$. When reducing the detector size, both, the heat capacity and conductance decrease. Because $C$ scales with the detector area and $G$ with the tether length, the response times get faster for trampolines with smaller detector size. The smallest trampolines performed best with response times of \SI{4}{\milli\second}.

\section{Discussion}
We demonstrate optimized trampoline structures made of SiN as enhanced NEMS-based infrared detectors. Compared to an earlier work on drum based nanomechanical resonators as infrared detectors~\cite{Piller2019}, which were also measured in this work as a reference, we could improve the NEP by two orders of magnitude with a minimum measured value of \SI{7}{\pico \watt \per \sqrt{\hertz}}. Further, we could also improve the response time by a factor three of the proposed infrared detector. The smaller the detector area, the longer become the tethers, which results in the enhance sensitivity. Larger detector areas can readily be obtained in a future design by increasing of the frame dimensions. With the current NEP, a detector area of $A=\SI{1}{mm}\times\SI{1}{mm}$ would result in the theoretical photon noise limit of $D^*\approx \SI{2e10}{\centi \meter \sqrt{\hertz} \per \watt}$ \cite{Rogalski2019}. Finally, the responsivity can be improved by using a transduction principle that does not require metal leads over the trampoline structure and hence is less deteriorating to the responsivity. Such NEMS resonators are promising thermal IR detector schemes with the potential to reach the ultimate photon noise sensitivity-limit. 
\\
\begin{acknowledgments} \label{sec:acknowledgements}
The authors wish to thank Sophia Ewert, Patrick Meyer, and Michael Buchholz for their support with the sample fabrication as well as Hendrik K\"ahler and Robert G. West for many fruitful discussions. We would also like to thank Georg Pfusterschmied for his support. This work is supported by the European Research Council under the European Unions Horizon 2020 research and innovation program (Grant Agreement-875518-NIRD). We further acknowledge funding from Invisible-Light Labs GmbH.
\end{acknowledgments}

\bibliography{literature}

\end{document}